\documentclass{sig-alternate-05-2015}
\usepackage[usenames, dvipsnames]{color}
\usepackage{eurosym}
\usepackage{enumitem}
\begin{document}

% Copyright
\setcopyright{acmcopyright}
%\setcopyright{acmlicensed}
%\setcopyright{rightsretained}
%\setcopyright{usgov}
%\setcopyright{usgovmixed}
%\setcopyright{cagov}
%\setcopyright{cagovmixed}

\newcommand\jg[1]{{\color{green}#1}}
\newcommand\tm[1]{{\color{blue}#1}}
\newcommand\ac[1]{{\color{black}#1}}

% DOI
%\doi{10.475/123_4}

% ISBN
%\isbn{123-4567-24-567/08/06}

%Conference
\CopyrightYear{2016}
\setcopyright{acmlicensed}
\conferenceinfo{BELIV '16,}{October 24 2016, Baltimore, MD, USA}
%\isbn{978-1-4503-4818-8/16/10}\acmPrice{\$15.00} %\doi{http://dx.doi.org/10.1145/2993901.2993911}
\isbn{tbd} \doi{tbd}
%
% --- Author Metadata here ---
%\conferenceinfo{BELIV'16}{October 24, 2016, Baltimore, MD, USA}
%\CopyrightYear{2007} % Allows default copyright year (20XX) to be over-ridden - IF NEED BE.
%\crdata{0-12345-67-8/90/01}  % Allows default copyright data (0-89791-88-6/97/05) to be over-ridden - IF NEED BE.
% --- End of Author Metadata ---

\title{On Regulatory and Organizational\\ Constraints in Visualization Design and Evaluation}
% \title{Alternate {\ttlit ACM} SIG Proceedings Paper in LaTeX
% Format\titlenote{(Produces the permission block, and
% copyright information). For use with
% SIG-ALTERNATE.CLS. Supported by ACM.}}
% \subtitle{[Extended Abstract]
% \titlenote{A full version of this paper is available as
% \textit{Author's Guide to Preparing ACM SIG Proceedings Using
% \LaTeX$2_\epsilon$\ and BibTeX} at
% \texttt{www.acm.org/eaddress.htm}}}

\numberofauthors{3} %  in this sample file, there are a *total*
\author{
% 1st. author
\alignauthor
Anamaria Crisan\\
       \affaddr{University of British Columbia}\\
       \affaddr{Vancouver, BC, Canada}\\
       \email{acrisan@cs.ubc.ca}
% 2nd. author
\alignauthor
Jennifer L. Gardy\\
       \affaddr{University of British Columbia}\\
        \affaddr{BC Centre for Disease Control}\\
       \affaddr{Vancouver, BC, Canada}\\
       \email{jennifer.gardy@bccdc.ca}
% 3rd. author
\alignauthor
Tamara Munzner\\
       \affaddr{University of British Columbia}\\
       \affaddr{Vancouver, BC, Canada}\\
       \email{tmm@cs.ubc.ca}
}
% There's nothing stopping you putting the seventh, eighth, etc.
% author on the opening page (as the 'third row') but we ask,
% for aesthetic reasons that you place these 'additional authors'
% in the \additional authors block, viz.
%\additionalauthors{Additional authors: John Smith (The Th{\o}rv{\"a}ld Group,
%email: {\texttt{jsmith@affiliation.org}}) and Julius P.~Kumquat
%(The Kumquat Consortium, email: {\texttt{jpkumquat@consortium.net}}).}
\date{24 October 2016}
% Just remember to make sure that the TOTAL number of authors
% is the number that will appear on the first page PLUS the
% number that will appear in the \additionalauthors section.

\maketitle
\begin{abstract}
Problem--based visualization research provides explicit guidance toward identifying and designing for the needs of users, but absent is more concrete guidance toward factors external to a user's needs that also \ac{have implications for} visualization design and evaluation. This lack of more explicit guidance can leave visualization researchers and practitioners vulnerable to unforeseen constraints \ac{%replaced a phrase here so we don't say external twice in two sentences
beyond the user's needs that can affect the} validity of evaluations, or even lead to the premature termination of a project.  Here we explore two types of external constraints in depth, regulatory and organizational constraints, and describe how these constraints impact visualization design and evaluation.  By borrowing from techniques in software development, project management, and visualization research we recommend strategies for identifying, mitigating, \textit{and} evaluating these external constraints through a design study methodology. Finally, we present an application of those recommendations in a healthcare case study.  We argue that by explicitly incorporating external constraints into visualization design and evaluation, researchers and practitioners can improve the utility and validity of their visualization solution and improve the likelihood of successful collaborations with industries where external constraints are more present.
\end{abstract}




%
%  Use this command to print the description
%

% We no longer use \terms command


\section{Introduction}
%Information visualization research can be stymied by factors external to the central problems the research is trying to address.
Simon's parable of \textit{The Ant on the Beach} asks readers to consider the trajectory of an ant as it walks along a beach:
``Viewed as a geometric figure, the ant's path is irregular, complex, hard to describe. But its complexity is really a complexity in the surface of the beach, not a complexity in the ant''~\cite{Simon1981}.
The parable highlights the importance of describing both the agent of action and the broader environment that acts upon that agent~\cite{Vicente1999}.

In problem--based visualization research and other user-centred methodologies, that agent is the user.
%and she forms the central tenet of user-centric design and evaluation models and methodologies.
%%TM: I find this wording very cryptic
%
While a focus on the user does not exclude consideration of her broader environment, little of the visualization research literature has been dedicated to precisely understanding how factors external to a user's needs affect design and evaluation~\cite{Lam2012}.
\ac{External factors can constrain the scope of the design space because, irrespective of user preferences, some solutions can never be implemented in their contextual environments. If researchers are unaware of these external factors from the project outset, they may develop and evaluate a visualization solution that cannot be used.}
%Consequently, visualization researchers can be caught unawares by external factors that can emerge throughout a project's life cycle.
%
For example, the authors of WeaVER, a tool that visualizes ensemble weather data, identified obstacles to data access, barriers of installing their visualization tool on locked--down workstations, and difficulty obtaining raw data %(sourced from institutional practices), I HAVE NO IDEA WHAT THIS MEANS
as factors affecting their ability evaluate the tool's design~\cite{Quinan2016}.

The discussion of external factors is not absent from the visualization research, \ac{but there does not exist more explicit guidelines toward incorporating factors from a user's contextual environment into visualization design and evaluation.}
%but these factors are typically considered to be secondary to a user's needs, and presented as obstacles, issues, or barriers that need to be worked around.
%
%
In this paper we propose that these external factors should be modelled as
 \textbf{constraints}~\cite{Vicente1999} that must be incorporated into visual and interaction design choices \ac{so as to yield relevant evaluations}.
We suggest strategies that visualization researchers can use to identify these constraints and provide recommendations for how constraints can be evaluated throughout a project's life cycle.
Finally, we demonstrate how our suggested strategies can be practically applied by presenting a case study in a healthcare environment, where external constraints can present many challenges for visualization researchers.

\section{Defining External Constraints} 
We have defined \textbf{external constraints} as any factor affecting visualization design and evaluation that is separate from the user's problem or needs \ac{and that are drawn from the user's contextual environment}. 
In this section, we further separate \ac{these} external constraints into two broad categories -- regulatory and organizational constraints.
In the context of this paper, we limit the definition of regulatory and organizational constraints to data access and the use of data for research purposes, because data is central to visualization research.

\textbf{Regulatory} constraints refer to legal requirements governing the collection, storage, and use of data.
In contrast, \textbf{organizational} constraints are policies and practices that are not necessarily encoded in law and that can vary across different institutions and across communities. 
Examples of organizational constraints can include policies around the protection of trade secrets, protectionist tendencies toward data, availability of financial resources, or institutional support for visualization projects~\cite{Carroll2014,Sedlmair2011}.
Importantly, organizational constraints encompass both the \textit{interpretation} and the \textit{enforcement} of regulatory constraints.
Differences of interpretation mean that different institutions can have different data access and use policies, some being more restrictive than others, while still conforming to the law.
Although these constraints are real, they should not discourage visualization researchers from collaborating with industries where regulatory and organizational constraints are present. 
By being aware of these constraints throughout the project's life cycle and explicitly incorporating them into visualization evaluation, researchers can enjoy fruitful collaborations, even within highly regulated industries. 

\subsection{\ac{Implications for Evaluation}}

\ac{Regulatory and organizational constraints have implications for design choices, often by restricting functionality and research processes (Section~\ref{subsec:agileBad} and~\ref{subsec:HypoGen})}.
\ac{As result, these external constraints provide additional parameters that need to be considered during evaluation or can define how evaluation should take place.}
\ac{For example, an additional parameter that needs to be evaluated is whether the visualization solution can be accessed by users, either by being installed on their work station or through web access, or whether IT constraints prevent local installations or uploading data to a web--based interactive platform.}
\ac{Such considerations can be missed when evaluating solely user's needs, as users themselves may not be fully aware of these constraints, or users may be inappropriately using their personal laptops for sensitive data and may not communicate they may be in violation of regulatory or organizational constraints.}

\ac{There are different consequences for failing to account for these constraints.}
%
%Most commonly, these constraints can affect project outcomes, for example the 
Failure to account for organizational constraints typically affects the \ac{validity of evaluations}, whereas failure to account for regulatory constraints may have \ac{legal} repercussions for a researcher \ac{and also the user}.
For example, ignored organizational constraints may result in project delays or termination, or a lack of adoption of the proposed solution.
%
%Litigation is unlikely, except in very extreme cases such as exposing trade secrets.  
%
However, researchers who fail to account for regulatory constraints are in violation of the law and could be subject to more severe penalties that involve the legal and judicial systems. 
\ac{It is thus necessary to evaluate that a project is in compliance with these external constraints throughout the project's life cycle.}
\subsection{Example: Hypothesis Generation\\Considered Harmful}\label{subsec:HypoGen}
One of the common arguments for the use of visualization is to facilitate new insights ~\cite{North2006}; that is, to generate new, testable hypotheses from data.
%
%Some visualizations specifically promote hypothesis generation by enabling data exploration, while tools indirectly support hypothesis generation by presenting data in a new way. 
%
However, in some highly regulated industries such as healthcare, finance, or the government, the ethics of exploring or mining data to generate new hypotheses is often controversial and is sometimes considered inappropriate or even illegal -- especially for data pertaining to individual people~\cite{Rindfleisch1997}.
%tweaked the block below a bit-jg
%

Both regulatory and organizational constraints influence exploratory analysis and hypothesis generation.
For example, organizations that routinely mine their users' data may have internal policies limiting who can mine this data, at what level of resolution (individual-level or aggregate), what can be reported and to whom, and what data may be unacceptable to use (for example, data from minors). 
%added some examples and the EU bit to the block below-jg
In highly regulated industries, legal boundaries also affect hypothesis--generating research. For example, personal data in Europe is subject to the recently adopted General Data Protection Regulation (EU 2016/679), which provides a framework governing multiple aspects of data use, including notice of collection, specified-purpose usage, consent, security, disclosure, access, and accountability.  Failing to adhere to the regulations can cost organizations fines of up to \euro 1,000,000.%i have no idea how to make the euro current symbol on this thing
%EUR1,000,000 in fines.
%
%deleted block below as you've already made this point
%Collaborating with highly-regulated industries can be challenging for visualization researchers as a number of constraints can arise.
%
%summarized the now-commented-out blocks below this with this single new paragraph

Visualization researchers who are new to highly-regulated environments might want to launch a visualization collaboration to specifically support hypothesis generation, yet might not be aware of the organizational or regulatory constraints that apply to their data that may preclude a successful outcome. Even researchers who have successful past collaborations with industrial partners with strict organizational constraints about the necessity of keeping proprietary data from leaking to the outside world may not realize the restrictions entailed by these kinds of regulatory constraints for any unauthorized data use whatsoever, even internally. 

%commented out as per aboveThe particular challenge of visualization research is that it generates hypotheses in two ways.
%
%commented out as per aboveThe first is through an agile development model that progressively elaborates upon the design.
%
%commented out as per aboveCollaborators may require a much crisper and definitive system design ahead of starting a research project, which is counter to visualization design and evaluation models and methods.
%
%commented out as per aboveThe second is that some visualizations tools are developed, often at the request of collaborators, to explicitly support hypothesis generation from data, but researchers may not be aware of restriction on hypothesis generation at a project's outset. 
%
%These constraints are not insurmountable, but do require active planning to identify and mitigate throughout the project life--cycle. 

\subsection{Example: Agile Development Considered Harmful}\label{subsec:agileBad}

Many visualization researchers advocate agile and iterative methods for visualization design and evaluation, but these approaches are often at odds with the rigid information technology infrastructure typically in place in institutions like hospitals, banks, or government agencies~\cite{Fitzgerald2013}. Moreover, concerns about the dangers of uncontrolled data exploration are frequently so central that they even extend to the realm of software development methods for tools to manipulate that data. Many organizations in highly--regulated industries remain firm in their use of waterfall software development models, despite their known problems and inefficiencies, rather than adopting more agile options~\cite{Larman2003}~\cite{Foster2013}. 

\begin{figure*}[t!]
    \centering
    \includegraphics[width=0.95\textwidth]{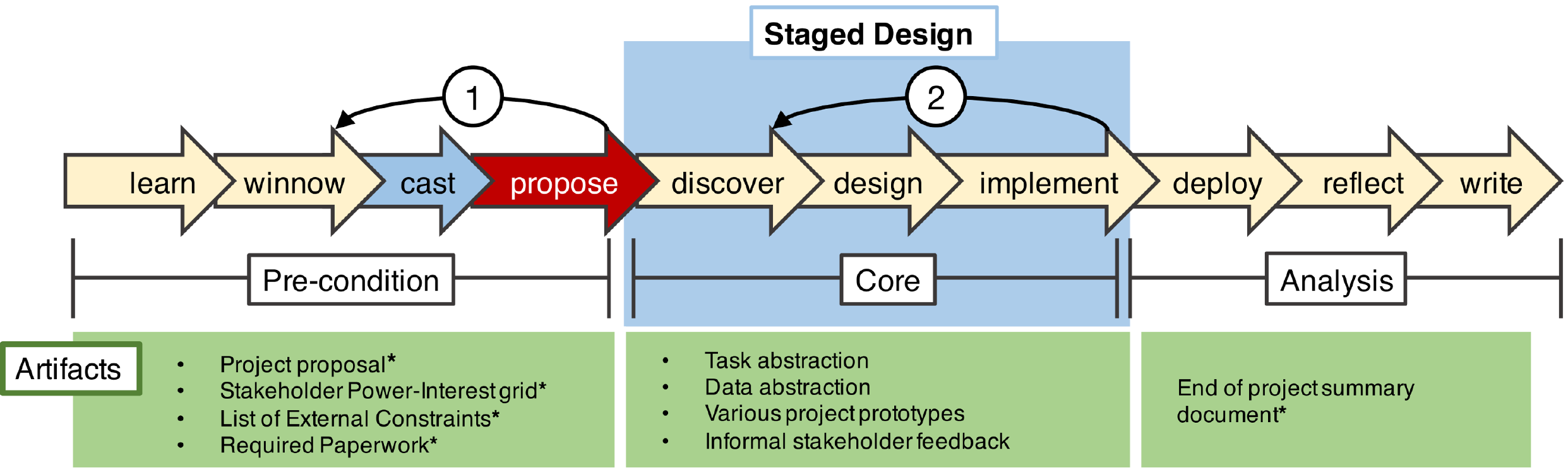}
    \caption{\textit{Summary of our proposed additions to the Design Study Methodology~\cite{Sedlmair2012}:} changes to the cast stage, a new propose stage, the generation of the starred artifacts, and identifying two of the many possible checkback cycles as required rather than optional.}
    \label{fig:dsmmod}
\end{figure*}

\section{Prior Work}
The central prior work appears both within the visualization literature and in other domains.
\ac{In this section, we appraise the extent to which prior work has equipped visualization researchers to identify, incorporate, and evaluate external constraints.}

%\noindent{\textbf{Visualization Methodologies.}}
\subsection{Visualization Methodologies}
The visualization research literature sets forth a number of models and methods to approach problem--driven design and evaluation projects~\cite{Meyer2013}.
These include contributions from our own group -- the Nested Model (NM) for Design and Validation~\cite{Munzner2009}, the follow-on Nested Blocks and Guidelines Model (NBGM)~\cite{Meyer2012,Meyer2013}, and a Design Study Methodology (DSM)~\cite{Sedlmair2012} -- and others, including Multi-dimensional In-depth Long-term Case studies (MILCs)~\cite{Shneiderman2006} and the Human-Centered Design Cycle~\cite{Lloyd2011}. %TM: very deliberate tactical choice to cite the original BELIV workshop paper here in addition to the followon IVJ article for the NBGM. will help address the question of relevance to this venue!
A central tenet of problem--driven research has been an emphasis on the needs of the target users and evaluating visualization design choices with respect to those needs.
\ac{The ``domain problem'' of the NM or the ``domain situation'' of the NBGM, and also more recent work by Winters ~\cite{Winters:2014} to further characterize domain situations via the NBGM through a new conceptual framework, \textit{could} be interpreted to include external constraints, but guidance is primarily offered toward identifying and evaluating user needs.} 
Similarly, the DSM and MILC approaches acknowledge the importance of considering the broader context in which visualization tools are deployed, but we argue they do do not sufficiently address external constraints. 

\ac{A small number of design studies and commentaries of design and evaluation methodologies have considered external constraints within the context of visualization research.} 
A study of large automotive companies warned of obstacles that are separate of ``technical challenges but [include] political or organizational requirements''~\cite{Sedlmair2010,Sedlmair2011}.
The authors suggested conducting pre-design studies to understand these factors in order to identify a feasible project path -- a sentiment that was shared in a position paper on pre-design empiricism~\cite{Brehmer2014}.
%
%I changed this because they really do actually say "test stuff throughout", or as MS called it "run multiple studies". it took a few reads to grasp formative and summative studies and separate design/eval studies for me, so that's why I revised this part. 
\ac{Both Brehmer~\cite{Brehmer2014} and Sedlmair~\cite{Sedlmair2010,Sedlmair2011} advocate for a variety of evaluation techniques at different design stages, with the thrust of their discussion focusing on a common agile motto ``test early, test often''.}
\ac{Another study by Lam~\cite{Lam2012} uses a \textit{scenario based} approach to evaluating visualization solutions that includes understanding environment and workplace practices, which they and others note is understudied in visualization research.}
\ac{Aside from identifying these evaluation scenarios through a literature review of visualization research, Lam \textit{et. al}~\cite{Lam2012} do not provide more detailed guidance towards the the types of external constraints or how they may be identified and evaluated.}
%

%the "common sense is not that common" point
\ac{The lack of explicit guidance toward evaluating visualization design with respect to external constraints means that individual researchers must devise strategies on an \textit{ad hoc} basis, which some researchers may be more successful at than others.}

%The consideration of external constraints is not completely absent from the visualization research literature. A study of large automotive companies warned of obstacles that are separate of ``technical challenges but [include] political or organizational requirements''~\cite{Sedlmair2010,Sedlmair2011}. %TM: very deliberate choice to include the original BELIV workshop paper here too. will help address the question of relevance!
%
%The authors suggested conducting pre-design studies to understand these factors in order to identify a feasible project path -- a sentiment that was shared in a position paper on pre-design empiricism~\cite{Brehmer2014}. %i am going to start saying pre-design empiricism so i can seem smarter-jg
%
%However, identifying and addressing all possible constraints at a project's outset is not realistic; these papers do not advise researchers how to deal with constraints that may arise or must be resolved later in a project's life cycle. 
\vspace{1mm}

%\noindent{\textbf{Design and Evaluation in Other Disciplines.}
\subsection{External Disciplines}
The design and evaluation of a system \ac{in the context of regulatory and organizational constraints} is not unique to the domain of visualization research or practice.
%edited the blocks below a bit to make the flow clearer to an agile n00b like me-jg

Some of the techniques used in visualization design studies are drawn from the larger set used in agile software development and related project management practices.
For many visualization research projects, applying the complete set of agile methodologies and practices may be inappropriate -- they do not capture some of the unique nuances of the visualization discipline and the agile framework can be too comprehensive and prescriptive for smaller, informal projects.
However, for large, formal collaborations in industries where the external constraints are much more pronounced, certain agile techniques from the software development literature can be useful.
%%said this block below already, and I don't really undertsnad what the last sentence means, so I have commented-out this block-jg
%In particular, many of these aforementioned disciplines with considerable constraints have serious concerns about agile methodologies because these disciplines require clearer design specifications at the project outset and wary toward agile methods~\cite{Larman2003}~\cite{Fitzgerald2013}; by  sharing a common agile philosophy, these concerns extend to visualization design and evaluation. 
%
In Section~\ref{subsect:methods}, we discuss specific techniques from the broader domain of agile software development that may be applicable \ac{toward design and evaluation of external constraints} for highly--regulated environments.

Cognitive Work Analysis (CWA) is another broad framework frequently deployed in developing technologies for the workplace, \ac{especially where regulations or safety are paramount considerations~\cite{Vicente1999}}. Its roots lie in systems thinking and ecological psychology, and it takes the most holistic view of a user and their contextual environment. 
A subset of CWA methods are frequently harnessed for visualization design and evaluation, particularly for task analysis.  
%added a bit here-jg
%
Importantly, CWA advocates undertaking a ``work domain analysis'' to understand a user's context because ``it imposes constraints on the actions of the actors''~\cite{Vicente1999}.

Collectively, the agile and CWA literatures offer a number of strategies for identifying and mitigating external constraints, but these strategies will be most useful only when appropriately contextualized for the visualization research domain.

%Also a thorough systematic review of communicable disease data visualization software, Carroll \textit{et. al.}~\cite{Carroll2014} identified two types of barriers that inhibited tool adoption: user, and system
%
%User barriers are defined as failing to meet a user's needs, and are familiar to visualization researchers.
%
%System-level barriers refer not to the visualization tool, but to the broader contextual environment in which that visualization tool is deployed.
%
%System level barriers include jurisdictional barriers, for example protectionist tendencies toward data sharing, resource (financial and personnel),technological, and lack of organizational support.
%

\section{Guidelines for Evaluating\\External Constraints}
We argue that the best way to mitigate external constraints is to proactively seek to identify them as early as possible, and to follow up by assessing whether they have been met as part of \textit{formative evaluation} efforts throughout the project's life cycle. 

We use the Design Study Methodology (DSM)~\cite{Sedlmair2012} as a scaffold to provide specific recommendations to visualization researchers.
We propose additional stakeholder roles within the cast stage and explicit communication strategies with them. We advocate the creation of several artifacts at many points, including at a new stage where a formal proposal is generated as part of a formative evaluation to assess project feasibility. These artifacts serve as checkbacks to specific previous stages, in contrast to the original DSM that simply encourages researchers to return to any prior stage of the framework as needs are noticed. We also argue for specific methods including a staged design process with generation of synthetic data as a stepping stone for access to the real data. Figure~\ref{fig:dsmmod} presents a summary of these recommendations. 
%it took me a minute to figure out that artifacts in Figure 1 refers to the outputs of the recommendations mentioned here. Maybe say something like "A summary of the artifacts arising from these recommendations..."?-jg

%
%this sentence below should be moved to the Figure 1 legend.
%Suggestions to modify existing DSM stages are indicated in blue, while new stages we propose are shown in red.
%
%Furthermore, although the original DSM encourages researchers to return to any prior stage of the framework as needed, we argue that some check-backs to earlier stages are critical to a project's success and should always happen.
%again, move this lil buddy to the Figure 1 legend
%-- we have emphasized those as numbered black arrows. 
%is how i reworded the block below correct?
%The most critical check-back occurs at the formative evaluation stage, when it must be assessed whether a project is feasible, given both user needs and the identified external constraints. 
%
%Finally, we propose producing additional artifacts in the pre-condition and analysis phases.
%move to figure 1 legend
%, which we've also indicated using an asterisk.
%
%
% ROLES
%
%
\subsection{Defining stakeholder roles}
\begin{figure}[t!]
    \centering
    \includegraphics[width=0.95\columnwidth]{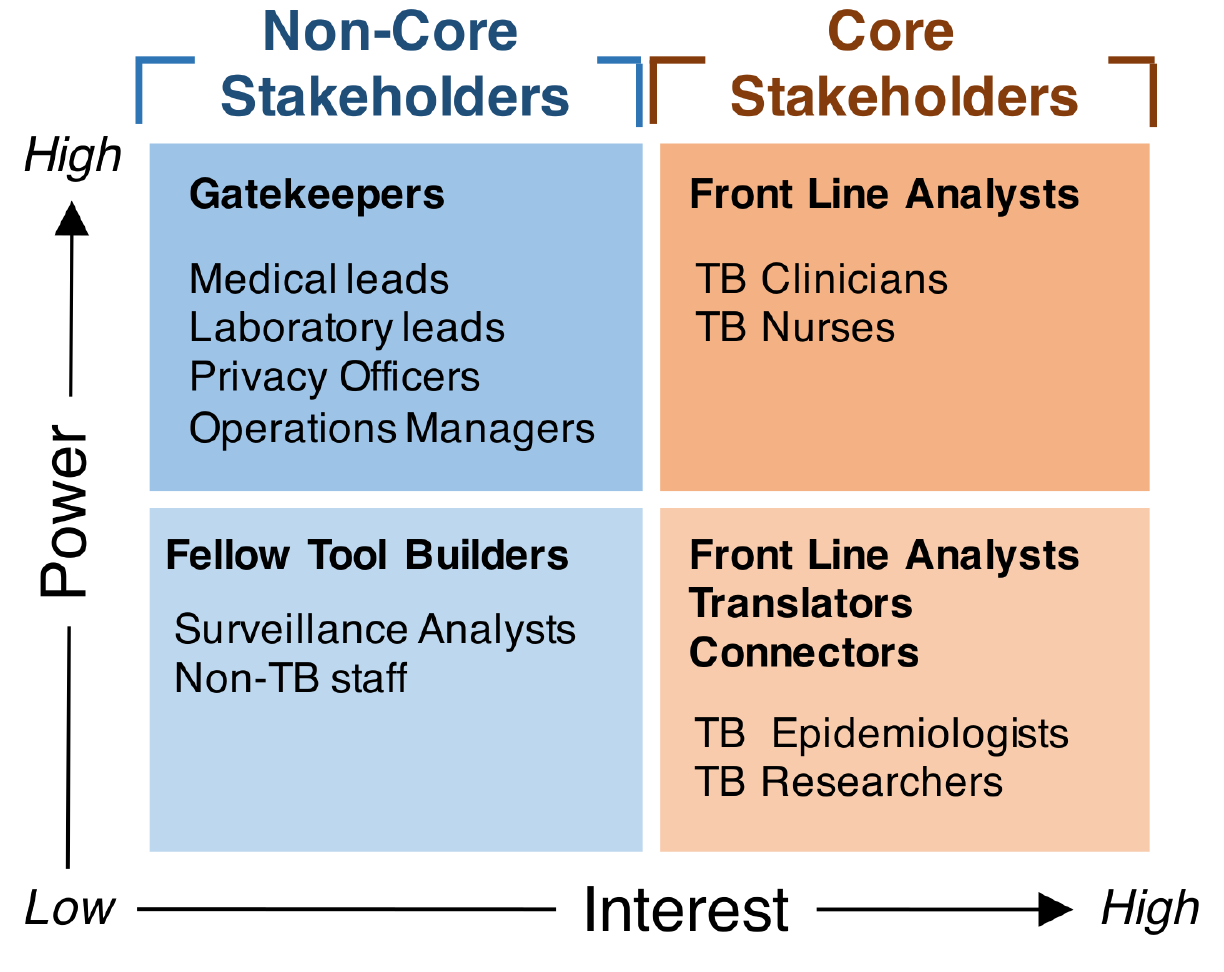}
    \caption[Power Interest Matrix]{\textit{Power Interest Matrix for identifying detailed roles during the cast phase.} Stakeholders are categorized into core and non-core groups according to their interest in project outcomes, and also as having low or high power. The specific roles identified in Section~\ref{casestudy} are included here as a concrete example.}
      \label{fig:stakeholders}
\end{figure}

The cast stage of the DSM pre-condition phase recommends that collaborators be cast as acting in one or more of several possible specific roles to help researchers identify the ways that relevant stakeholders might become involved in a project: front-line analyst, gatekeeper, translator, connector, or fellow tool-builder.

\vspace{1mm}

\noindent{\textit{\textbf{Recommendation 1: Classify stakeholders according to power over and interest in project outcomes.}}} 
We argue that this classification should be extended to further improve stakeholder identification and management: these roles should be further stratified according to the amount of power over and interest in the project outcomes for each stakeholder. 
Using a power-interest grid can help identify stakeholders~\cite{Mendelow1991} -- particularly gatekeepers -- that may not be immediately obvious; for example, individuals who are not directly involved in a project but who can affect the project through their role in assuring compliance with regulatory or organizational constraints. 
Stakeholders that have high interest in a project's outcomes, whether low- or high-power, typically form a core group with whom researchers closely collaborate; these core stakeholders will be actively involved in visualization design and evaluation (\ac{both formative and summative}) and they also supply the motivation and needs for a visualization solution.  
Indeed, a visualization project may be initiated through these high-interest stakeholders. 
\ac{Here, we do not prescribe a specific type of formative evaluation methodology, but note that much of the evaluation studies proposed in visualization research, including interviews, questionnaires, think-out-loud, and laboratory experiments, are targeted toward these core stakeholders.}
Non-core stakeholders are those with whom researchers do not collaborate directly and who are thus classified as low-interest.
\ac{Often, stakeholders with high power, but low interest in project outcomes are those that must be consulted with in order to access data and get approval to conduct the research; the DSM classifies these stakeholders as Gatekeepers.}
Gatekeepers can be individuals that oversee the appropriate access and use of data, both at the outset and throughout a project, or an institutional review board that provides initial approval for data access and use. 
While this quadrant of high--power, low--interest stakeholders are unlikely to participate in visualization design processes, individual gatekeepers (but not entire review boards) should be included in \ac{\textit{at least} guideline checking formative evaluations~\cite{Andrews2008},} to confirm compliance with regulatory and organizational constraints.
\ac{Finally, there are stakeholders with low interest and low power in visualization project outcomes.} These individuals may have an intellectual interest in project outcomes, such as other researchers building analytical tools; while these individuals will not take part in either \ac{design or evaluation} they may form useful allies in the institution and inform researchers about external constraints. 
\vspace{1mm}

\noindent{\textit{\textbf{Recommendation 2: Actively manage communication with stakeholders.}}} 
While DSM does indicate that poor rapport can be a potential pitfall to a project's success (PF-9)~\cite{Sedlmair2012}, it does not provide explicit guidance towards managing communication with stakeholders. 
Ineffectively managing stakeholder communications can impact the discovery of regulatory or organizational constraints, \ac{which in turn impacts the validity of evaluations and could even lead to premature termination of the project}.
\ac{Good communication with stakeholders is also critical for carrying out formative evaluations with core stakeholders and guideline checking evaluations with gatekeepers of prototypes developed through the staged design process (Section~\ref{subsect:methods})}. 

We recommend using the power-interest grid of Recommendation 1 as the framework for managing stakeholder communications. For core stakeholders, communication can be informal and will be more frequent than with non-core stakeholders. 
For non-core stakeholders with high power over a project's outcomes, we recommend more formal communication.
Some institutions will already have polices in place for communication templates and the timeliness of those communications, but when such guidelines are not available, we recommend a formal, plain-language brief that is distributed to these stakeholders.
These briefs may be more frequent at the beginning of the project, especially if there is uncertainty around the nature of agile development methods and the design study framework, and may become less frequent over time. 
\ac{Communication briefs should emphasize the findings of evaluations carried out during the design process.}

Effective communication with stakeholders can have the added benefit of improving institutional awareness of visualization research, which may make future projects easier to conduct.
%
%
% Artifacts
%
%
\subsection{Generation of Additional Artifacts}
Conducting  pre-design studies~\cite{Sedlmair2012}~\cite{Brehmer2014} to assess a project's feasibility and to identify regulatory and organizational constraints is important.
%
%These pre-design studies may happen ahead of the pre-condition of the DSM, or may occur throughout.
%TM hmm. I think it's a confusing point. We should talk if you think it's crucial to make somehow. 
%
In its original form, the DSM recommends going directly from the cast stage to the discover stage of the core phase, but we argue that this transition is premature and recommend an explicit propose stage between the two.
This new propose stage entails creating additional project artifacts that help to guide formative evaluations of user needs, in addition to identifying regulatory and organizational constraints.
These artifacts are in addition to the task and data abstractions and the prototypes that already form part of the DSM's core phase. 
\vspace{1mm}

\noindent{\textit{\textbf{Recommendation 3: Create a formal proposal document.}}}
The most important of these artifacts is a project proposal that summarizes the evidence gathered in the pre-design studies \ac{and consultation with high--power stakeholders} into a single document.
This proposal document should be assessed by both researchers, core stakeholders, and Gatekeepers, before proceeding to the core phase of the DSM.
\ac{Throughout various stages of formative evaluation during the design process, this document can serve as the basis for the guideline checking that will be carried out with Gatekeepers~\cite{Andrews2008}.}

%changed the first sentence of this cause in university bureaucracy land, everything needs a dang proposal, even non-research projects. Gah. -jg
%---shortened some stuff here to get back some space, I think using the Andrew's article may be more effective because it's vis oriented --
%Many of the service-oriented activities engaged in by institutions like hospitals or government agencies do not require formal proposals, whereas research activities are usually conducted in academic settings that are rife with such documents. 
%When engaging in collaborative research with institutions that have significant regulatory and organizational constraints, we strongly recommend creating a formal proposal before proceeding to prototype development, even if this document is not required by the institution's governance model.
%
Institutions may have specified proposal templates but if a proposal template does not exist, we recommend communicating --at minimum -- the project's scope, including user needs, known external constraints, data requirements and uses, who is involved and what they will be doing, and a brief description of the design process and evaluation procedures.
This proposal will typically be refined through a process of discussion with stakeholders. 
Although the DSM encourages researchers to backtrack to any of the proceeding steps without requiring any checkback loop explicitly, our extension proposes that the completion of the final proposal document should trigger a \textit{required} revisition of the winnow stage, as shown by  arrow 1 in Figure~\ref{fig:dsmmod}. The goal is to evaluate whether the project can be completed in a timely manner and is mutually beneficial to stakeholders and researchers.
\vspace{1mm}

\noindent{\textit{\textbf{Recommendation 4: Create a summary document at the end of a project.}}}
At the end of the project we recommend creating a summary document that expresses -- in plain language -- the ways in which the project addressed a relevant domain problem in light of external constraints.
The project conclusion document is meant to complement the initial project proposal by highlighting the resulting mutual benefits of the project for both researchers and stakeholders. A research paper describing the project outcomes in terms suitable for an academic audience of other researchers who grapple with visualization design and evaluation issues is not a suitable stand--in for this document, which is aimed at a very different audience with different concerns. In some cases, the process of abstraction that was undertaken  by the visualization researcher needs to be inverted so that the solution can be described in domain-specific terms in a way that makes sense to the intended audience. 
However, this conclusion document can be helpful for educating stakeholders on the processes and relevance of visualization research~\cite{Lloyd2011}, \ac{especially if the document emphasizes how the results of various evaluation studies are in line with individual stakeholder needs and also institutional policies}. It has the potential of laying out important groundwork so that future visualization research projects are easier to conduct. 
%
%
% Methods
%
%
\subsection{Methods}\label{subsect:methods}
Once researchers and stakeholders have an understanding of users needs as well as external constraints, both should be integrated into the visualization design and evaluation process.

\vspace{1mm}

\noindent{\textit{\textbf{Recommendation 5: Use a staged design process.}}}
The staged design model~\cite{McLachlan2008} proposes incremental prototype development through a series of stages, making it possible to progressively gain access to users and resources that may not be accessible at a project's outset and to accommodate changes in the stakeholders' context and environment that arise over a project's life cycle.
Each stage consists of requirements--gathering and prototype development to produce a minimum viable product with progressively improving fidelity. The model  should conclude with a formal evaluation that specifically demonstrates whether the tool and development process is in compliance with regulatory and organizational constraints, in addition to meeting stakeholder needs.

%
%The staged design approach fits with the agile mantra of ``test early and test often''.
%
At the end of each design stage, we highly recommend that researchers and collaborators explicitly evaluate together whether or not it is feasible to proceed to the next stage of development, as shown by arrow 2 in Figure~\ref{fig:dsmmod}. By proactively checking on feasibility in this way, initially unforeseen constraints that arise later in the project are surfaced as early as possible, to minimize later adverse impact on researchers such as a loss of access to data or people. 

%rather than 
%researchers may be less affected by 
%affect their ability to access certain data or stakeholders or continue with the project. 
%
Using a staged design process also allows researchers to plan and prioritize minimal viable products, some of which may be valid visualization research contributions in themselves -- even if a project is terminated ahead of the originally planned schedule.
\vspace{1mm}

\noindent{\textit{\textbf{Recommendation 6: Use synthetic data early on if real data is not immediately available.}}} 
As discussed in Section~\ref{subsec:HypoGen}, some industries have concerns around hypothesis-generating research related to both the agile design process and the types of insights that can and should be drawn from data.  
Stakeholders in these industries may \textit{want} visualization tools to support hypothesis generation \ac{of individual level data}, but nevertheless may wish to impose limits on the types of uncontrolled exploration a user can conduct~\cite{Rindfleisch1997}.
At a project's outset, it may not be clear  yet how to operationalize such limits, which puts researchers and stakeholders in the difficult position of potentially violating regulatory constraints.
These constraints can make collaborators wary of sharing real data at a project's outset, thus impeding the launch of a potential collaboration with visualization researchers.
One way to overcome this constraint is to use synthetic data in early design stages and gradually earn the trust necessary to gain access to real data in later stages.
Synthetic data is never a perfect substitute for real data because it lacks nuances that may be of interest of stakeholders; consequently, the use of synthetic data affects the validity of evaluations of a prototype's utility.
For example, synthetic data is often very clean, avoiding the problems of missing or erroneously entered data that are often present in real data; while such noise can be simulated, the scope of possible errors may be difficult to fully understand and incorporate in synthetic data generation.
The nuances of supporting users in handling dirty data might therefore be absent from a design process and evaluation process where only clean data is used. 

In spite of these limitations, synthetic data can nevertheless an effective means to demonstrate a tool's functionality and to allow researchers and stakeholders to have concrete discussions about what aspects of functionality should be limited. 
By graduating from synthetic to real data and modifying the rigor of evaluations over time, what may be lost in initial evaluation validity can be gained in collaborators' trust. Starting with synthetic data can be a viable alternative to giving up on the project during early stages due to initial regulatory and organizational constraints. 

\section{Case Study: Healthcare}
\label{casestudy}
\ac{In this section we provide a concrete example of how to interpret our recommendations through a case study in healthcare, in an approach similar to Winters \textit{et al}~\cite{Winters:2014}. Case studies provide an opportunity to dive deep into a specific domains to provide insights into a phenomenon that may be transferrable to other domains~\cite{Flyvbjerg2006}, and their benefits for visualization research has been argued by Shneiderman and Plaisant in their ethnographically informed proposal for multi-dimensional in-depth long-term case studies (MILCs)~\cite{Shneiderman2006}.}  %winters paper was published in BELIV 2014.
%TM: i don't know what contextualize means here. think we need to say something more clear.
%AC : I like the way you've put it here, I think it's clearer.

Healthcare systems comprise two disciplines -- clinical medicine and public health -- that must work together to improve the health of both individuals and populations.
Public health focuses on prevention and control activities, while clinical medicine focuses on diagnosis and treatment~\cite{THCSPH2016}.
While clinical medicine tends to be the domain of specialist health care providers such as clinicians, nurses, and pharmacists, public health professionals are more diverse. In addition to the aforementioned providers, their roles include, but are not limited to, epidemiologists, statisticians, researchers, politicians, and other community leaders.

In some cases these two disciplines can operate nearly independently of one another, but in others they must work more closely together to deliver patient care.
The world of communicable disease prevention and control is an example of the latter, where disciplines must share knowledge and make decisions together -- clinicians guide the management of individual patients with a disease, while public health authorities manage the disease at a population level.
Although they must work together, the different traditions informing public health and clinical medicine mean that there is often a knowledge translation gap, where the knowledge and data generated by each discipline is siloed,  ultimately affecting the ability of these disciplines to work together~\cite{Woolf2008}. 

Visualization tools can help stakeholders in public health and clinical medicine to more readily share knowledge and insights that support decision making at patient and population levels. 
But in order to be most effective, visualization researchers need to operate within the bounds of the significant regulatory constraints that apply to healthcare and healthcare data, as well as the organizational constraints in healthcare, which can differ between public health and clinical medicine. 
%

%#############################################################
%					SUBSECTION Case Study : Constraints
%#############################################################
%
\subsection{Constraints in Healthcare}
%If visualization researchers can only react to those regulatory constraints as they are discovered, then in the worst case these researchers risk breaking the law by inappropriately using health data, or in more common case can risk their project being shut down by regulators.

%#############################################################
%					SUBSUBSECTION Case Study : Reg Constraints
%#############################################################
%
\noindent{\textbf{Regulatory Constraints.}}
The law distinguishes between primary and secondary use of health data~\cite{Safran2007}. 
Primary uses of health data are those associated with the direct and immediate care of a patient, while secondary uses are all other uses that do not directly contribute to a patient's care. This category includes all research using health data.
While the law does not prevent the secondary use of health data, it does place restrictions on such usage that are meant to balance an individual's right to privacy and confidentiality while simultaneously stimulating progress in public health and clinical medicine.
Oversight and implementation of these regulatory constraints is not consistent across different institutions~\cite{Safran2007}.
\vspace{-1.5mm}

\noindent{\textbf{Organizational Constraints.}}
It is recognized that the secondary use of health data is a ubiquitous and necessary practice, but data access models vary considerably and are not transparent, which affects research productivity~\cite{Safran2007}.  
Many institutions are wary of uncontrolled secondary use of data~\cite{Rindfleisch1997}, in which any researcher can explore any manner of hypothesis in a dataset without clear benefit to the patient.
While exploratory hypothesis-generating research is important, it is a hotly debated as a practice because it is ultimately the patient, and not the researcher, that bears the full burden of accidental data disclosure. 

Researchers who request access to health data are often required to have a well-formed hypothesis at the project outset, in addition to outlining their analytical methods.
As was discussed in Section~\ref{subsec:HypoGen}, these restrictions on hypothesis-generating research affect not only the functionality of data visualization tools, but also the application of agile-like methods for developing them.  

Aside from organizational practices that enforce regulatory constraints, there also exist hierarchical and political structures that can result in protectionist tendencies toward data.
These protectionist tendencies can arise because a particular individual is responsible for stewarding the appropriate use and interpretation of health data, or because researchers are hesitant to share data that was costly and time-consuming to obtain. 

%
%#############################################################
%					SUBSECTION Case Study : Applying Revised Methods
%#############################################################
%

\subsection{Lessons Learned in Developing a TB\\Decision Support Tool}

Our proposals for integrating constraints into the visualization design and evaluation grew out of a specific project in a highly-regulated healthcare domain. 
\vspace{1mm}

\noindent{\textbf{Application: Tuberculosis Prevention and Control.}}
Of the many communicable diseases managed by a public health agency, tuberculosis (TB) is one of the most interesting. It has a long history of infecting humans, with TB found in the remains of mummies and tales of ``consumption'' a popular theme within popular culture~\cite{Daniel2006}.
Despite this long history, medicine has not yet succeeded in eliminating TB. In 2012 alone, there were 8.6 million new cases of symptomatic TB and 1.3 million deaths, and as much as 1/3 of the world's population is thought to be infected with a latent, asymptomatic form of the disease~\cite{WorldHealthOrganization2013}.
New strategies to manage existing cases and prevent future ones are clearly needed. Opportunities for designing and delivering new interventions to combat TB are available through exploring and mining patient-level data in electronic health records, population-level data in disease registries, and even molecular data describing pathogenic microbes rather than human individuals. 
\vspace{1mm}

\noindent{\textbf{Collaboration Context.}}
We report and reflect upon a collaboration with stakeholders involved in TB prevention and control at the British Columbia Center for Disease Control (BCCDC). Our goal was to build a decision support tool to facilitate our users' routine workflows and to allow exploratory analysis in support of new intervention development.
We did not set out to construct a fail--safe healthcare application; rather, we set out to collaboratively explore how visualization of our stakeholders' data could support decision making.
At the start of our collaboration, armed solely with existing visualization design guidelines, we were often reacting to previously unknown regulatory and organizational constraints rather than proactively mitigating them -- and at one point faced the risk that the project would not move forward.

At the outset of our collaboration, we engaged with a small group of stakeholders at the BCCDC that consisted of clinicians, nurses, epidemiologists, and researchers. 
This group had worked together extensively in the past, and had a history of productive prior research collaborations. 
We engaged in discussions with them about a project that explored the utility of data visualization to provide multiple perspectives on the spread of TB through the province of British Columbia over time.
The insights this group of stakeholders would gain from the tool would help inform future policies and practices in TB prevention and control.
Our discussions around the project and its objectives were informal, and the data we had intended to use for tool development had received prior approval for research use.
With a promising collaboration on our hands, we began to engage in discussion with these stakeholders about the data types in use at BCCDC and the ways our stakeholders used these data for both routine and high-level policy decision-making.
\vspace{1mm}

\noindent{\textbf{Discovering Lurking Constraints.}}
While we were focused on assessing our stakeholders' needs and their primary research question, we confronted the first regulatory and organizational constraints that would temporarily suspend our project's progress.
Over the course of our project, the BCCDC had changed the way it gathers Public Health data, and how use of this data for research was to be governed.
%The Integrated Public Health Information System (iPHIS) database that had previously warehoused the many public health data streams collected in BC was being replaced by a completely new platform, Panorama. 
%As a result of updating a Public Health data collection technology platform, the way in which the secondary use of data was approved  was changed. -- SAVE SPACE BYT SHORTENING THIS
%
Not only were data approval polices changing, but so too were the individuals responsible for the approvals \ac{(referred to internally as data stewards)}.
As part of taking on their mandate, the new TB data steward took stock of current research projects, and flagged our visualization project for re-assessment.
His concern was that the project did not clearly outline how it may be directly beneficial to patients and so had the potential to be deemed unethical.
Although an ethics committee had reviewed and approved the use of our data for secondary purposes, the new data steward indicated that we needed to provide a more detailed justification for our specific project before we could continue.
\vspace{1mm}

\noindent{\textbf{Identifying Additional Gatekeepers.}}
Neither we nor our collaborators had anticipated this intervention by the data steward.
As we began to gather information about the necessary next steps to take in order to continue our project, we  sought to understand other aspects of the organizational structure and identify other gatekeepers that might further impede our project's progress.
We took on the exercise of creating a power-interest grid (\textit{Recommendation 1}) and over time we stratified our TB stakeholder group as follows:
\begin{itemize}[leftmargin=0.1in]
\setlength\itemsep{0.05 mm}
\item \textbf{High Interest}, \textbf{High Power} \textit{Front-line Analysts (TB clinicians and nurses)}: Data for individual patients was primarily and controlled by and accessed through clinicians and nurses. With a strong interest in using data to develop new policy and practice, these individuals formed part of our core stakeholder group.
\item \textbf{Low Interest}, \textbf{High Power} \textit{Gatekeepers (Departmental Medical Leads, Laboratory Leads, Privacy Officers, and Operations Managers):} Both medical and laboratory leads must sign off on data usage, though they may not be directly involved in TB control or invested in our project outcomes. Privacy officers and operations managers also enforce regulatory processes. One particularly powerful, but difficult to reach, stakeholder was the organization's IT department, as they controlled the users' workstations and permissions for software installation. These individuals did not form part of our core stakeholder group. 
\item \textbf{High Interest}, \textbf{Low Power} \textit{Front-line Analysts and Connectors (TB epidemiologists and researchers)}: In our study, researchers had control over the use of the pathogen-level molecular data they had generated and epidemologists could advise us on the use of patient-level case data, but neither class of stakeholder had the authority to sign off on data usage beyond the molecular data. Still, as integral parts of the TB control team, they were interested in our project outcomes, and were part of our core stakeholder group.
\item \textbf{Low Interest}, \textbf{Low Power} \textit{Fellow Tool Developers (Non-TB analysts)}: Other groups around the BCCDC were interested in visual analytic tools for their own applications outside of TB, but were outside of our core stakeholder group.
\end{itemize}

We established a rough communication plan (\textit{Recommendation 2}) to engage with these stakeholders in order to proactively identify important constraints moving forward.
Often our communications were one-on-one discussions, but when availability afforded it, we conducted large group meetings with both core and non--core stakeholders. 
\vspace{1mm}

\noindent{\textbf{Finding Constraint Impact on Functionality.}}
As we identified different stakeholders, we learned of more organizational constraints that would affect the functionality of the decision support tool we intended to build.
We learned that our tool should not support what might, at first glance, seem to be obviously useful data wrangling functionality such as merging multiple datasets, correcting data errors, or entering missing data.
There were institutional policies in place that governed how and by whom multiple datasets could be merged because of concerns around privacy -- as more datasets are linked together, there is a higher likelihood of potentially re-identifying patients. 
%
%dont' really need the block below so I commented it out -jg
%BCCDC policies would assess the risk a re-identification, and make a consideration about the ethical use of such data. 
%
Furthermore, institutional procedures were also in place to correct errors or handle missing data in a systematic way, and again were carried out only by select individuals.

Given the constraints that precluded data wrangling, we recognized that our tool would function best as a data viewer that could alert core stakeholders to missing or incorrect data but not permit them to change the underlying dataset.
Furthermore, our tool needed to flexibly handle whatever data and data types different core-stakeholders were permitted to access, ranging from clinicians and nurses allowed to access individual patient data, to generalist users who should only be shown aggregate data.
\ac{These additional requirements affected functional requirements and served to constrain our design space.}

\vspace{1mm}
\noindent{\textbf{Finding Constraint Impact on Real Data Access.}}
Some stakeholders unfamiliar with the design process considered it odd that we had not already established the visual and interaction design choices for our decision support tool. They also found it unusual that we intended to conduct a research project to figure out what those design choices should be.
Thus, our research was initially perceived by some as an uncontrolled use of secondary data (Section ~\ref{subsec:HypoGen}), and several Gatekeepers were unwilling to allow us to use real data at the outset. 
We thus considered at length how to develop a strategy that would gain these users' trust in our research methods. 

\vspace{1mm}
\noindent{\textbf{Finding Constraint Impact on Tool Integration.}}
Through several stakeholders, we also learned about the impact of the information technology (IT) group's policy that workstation environments should be locked down.
A lengthy approval process was required to install new software or host custom web applications on institutional servers. 
Accessing web applications for data analysis was also prohibited because data could not leave institutional servers.
%moved the two blocks below up since i think they fit a little better here -jg
Part of the reason for these constraints is that the  IT group manages workstations in many healthcare settings, including not only research workstations but also those used in clinical care, resulting in very restrictive workstation policies. 
%
%Restricting the types of applications that can be downloaded and installed on hospital workstations as well as administrative access for these computers is clearly important for patient privacy and security, but difficulties arise when restrictive policies appropriate for clinical practice are applied to all workstation users in a healthcare system, even those used by researchers.  

%
One tool that could be used in the existing constrained environment -- and indeed was widely used by BCCDC epidemiologists -- was R.
Although the version of R available on workstations was outdated and could only be updated by IT, we knew there were plans to update it, and decided that a R--based tool would be a viable implementation solution that fit into BCCDC's existing organizational infrastructure.
\vspace{-2mm}

\noindent{\textbf{Changing Strategies for Emerging Constraints.}}
The identification of these constraints and our assessment of their impacts on our decision support tool's functionality, utility, and stakeholder adoption allowed us to reformulate our project's trajectory.
We prepared a project proposal for our core-stakeholders and gatekeepers that outlined clearer objectives for our tool in light of the various regulatory and organizational constraints we identified (\textit{Recommendation 3}).
Importantly, we also indicated how stakeholders would be involved in evaluating our compliance with these constraints. 
\vspace{-2mm}

\noindent{\textbf{Building Trust Through Staged Design.}}
We planned for a staged design process based upon different datasets (\textit{Recommendation 5}).
The first stage of design would use only data available in routinely collected administrative datasets, while later stages would combine this data with laboratory and contact network (who was exposed to an infectious individual) datasets. 
In this way, we would produce minimum viable products for the most commonly used dataset first, and less commonly used datasets later.
Although we could not use the real data, we had access to the structure and aggregate statistics of the real datasets because they were made public through BCCDC's annual reports.
As much as we were able to, we based our synthetic datasets off of the real data (\textit{Recommendation 6}).
We hypothesized that if stakeholders were enthusiastic about how the decision support tool could visualize their most commonly used dataset, albeit as demonstrated by synthetic data, that this demonstration may encourage them to move toward using the tool with real data.

We conducted focus groups and developed paper prototypes during the first design stage to gather user requirements and marry those to known regulatory and organizational constraints. 
The inability to install our tool on stakeholder workstations led us to rely on chauffeured demos~\cite{Lloyd2011}, using a workstation with a more current version of R, to conduct evaluations at the conclusion of the design stage. 
We gathered qualitative evaluations of the tool's perceived utility and the validity of our design choices. 
To evaluate compliance with regulatory and organizational constraints, we worked closely with BCCDC's privacy officer (Guideline checking evaluations).

Although not rigorous, our evaluation gave stakeholders an opportunity to see what a decision support tool that visualizes TB data could do and how it could help them.
Furthermore, instead of discussing abstract notions of how this tool may or may not be beneficial to patients in the long term, we could engage in more concrete discussion with stakeholders -- especially Gatekeepers -- about what functionality was appropriate and what was not. 
We summarized the design and evaluation progress, highlights, and outcomes of our collaboration at a larger group meeting following the conclusion of the first design stage.
We emphasized how a visualization tool could responsibly incorporate regulatory and organizational constraints that are meant to safeguard patient data, \ac{and demonstrated this capacity by emphasizing the results of formative evaluations that various stakeholders had participated in.}
The success of this initial stage has initiated concrete discussion by both core--stakeholders and gatekeepers toward evaluating the tool using real data.
Thus, what could have been a failed start due to unforeseen initial constraints has evolved into a viable project with organizational support for its continuation.

\section{Conclusions}
We have put forth that rather than being peripheral considerations to the visualization research, external constraints are an important component of visualization design and evaluation.
The current visualization research literature acknowledges that these constraints are present, but does not offer explicit guidance around their identification or their incorporation into the design process.
By using strategies drawn from software development, project management, and visualization research, we have provided several recommendations that modify an existing Design Study Methodology such that researchers can proactively identify and evaluate potential constraints throughout a project's life cycle.
\ac{Although these strategies may not be applicable for all visualization design projects}, they are beneficial when visualization researchers set out to engage in collaborations with industries that are highly regulated, or with large institutions with complex lurking bureaucracies. 
\ac{Furthermore, we encourage visualization researchers to draw from knowledge in external disciplines, like agile software development, project management, and cognitive work analysis, to supplement design and evaluation processes in their projects.}
If visualization researchers are not aware of the complex backdrop against which they conduct their research, they risk being discouraged and may not apply their talents where they are sorely needed.
By explicitly evaluating a visualization tool with respect to regulatory and organizational constraints, researchers can increase the likelihood of success of both their project and their collaboration.

%\end{document}  % This is where a 'short' article might terminate

%ACKNOWLEDGMENTS are optional
\section{Acknowledgments}
We wish thank the following members of the BCCDC: Dr. James Johnston, Dr. Maureen Mayhew, Dr. Victoria Cook, Nash Dahlla, Dr. Jason Wong, Dr. James Brooks, Johnathan Spence, Laura MacDougall, Michael Coss, Ciaran Aiken, and David Roth. We would also like to thank the UBC CS visualization group for their helpful feedback: Matthew Brehmer, Madison Elliott, Zipeng Liu, Dylan Dong, and Kimberly Dextras--Romagnino. Anamaria Crisan is funded through a CIHR-Vanier Scholarship.

\bibliographystyle{abbrv}
\bibliography{references}  % sigproc.bib is the name of the Bibliography in this case

\begin{thebibliography}{10}

\bibitem{Andrews2008}
K.~Andrews.
\newblock {Evaluation Comes in Many Guises}.
\newblock {\em BELIV'08Workshop, CHI 2008}, pages 8--10, 2008.

\bibitem{Brehmer2014}
M.~Brehmer, S.~Carpendale, B.~Lee, and M.~Tory.
\newblock {Pre-design empiricism for information visualization: scenarios,
  methods, and challenges}.
\newblock In {\em {Proc. ACM BELIV Workshop}}, pages 147--151, 2014.

\bibitem{Carroll2014}
L.~N. Carroll, A.~P. Au, L.~T. Detwiler, T.-C. Fu, I.~S. Painter, and N.~F.
  Abernethy.
\newblock {Visualization and analytics tools for infectious disease
  epidemiology: A systematic review.}
\newblock {\em Journal of biomedical informatics}, 51:287--298, apr 2014.

\bibitem{Daniel2006}
T.~M. Daniel.
\newblock {The history of tuberculosis}.
\newblock {\em Respiratory Medicine}, 100(11):1862--1870, 2006.

\bibitem{Fitzgerald2013}
B.~Fitzgerald, K.~J. Stol, R.~O'Sullivan, and D.~O'Brien.
\newblock {Scaling agile methods to regulated environments: An industry case
  study}.
\newblock {\em International Conference on Software Engineering}, pages
  863--872, 2013.

\bibitem{Flyvbjerg2006}
B.~Flyvbjerg.
\newblock {Five Misunderstandings About Case-Study Research}.
\newblock {\em Qualitative Inquiry}, 12(2):219--245, apr 2006.

\bibitem{Foster2013}
R.~Foster.
\newblock {Don't Go Chasing Waterfalls: A more Agile Healthcare.gov}.
\newblock {\em The Newyorker}, 2013.

\bibitem{THCSPH2016}
{Harvard T.H. Chan School of Public Health}.
\newblock {Distinctions Between Public Health and Clinical Medicine}.
\newblock \url{http://www.hsph.harvard.edu/about/public-health-medicine/}.
\newblock Accessed: 2016-03-21.

\bibitem{Lam2012}
H.~Lam, E.~Bertini, P.~Isenberg, C.~Plaisant, and S.~Carpendale.
\newblock {Empirical studies in information visualization: Seven scenarios}.
\newblock {\em IEEE Trans. Visualization and Computer Graphics (TVCG)},
  18(9):1520--1536, 2012.

\bibitem{Larman2003}
C.~Larman and V.~Basili.
\newblock {Iterative and Incremental Developments: A Brief History}.
\newblock {\em Computer}, 36(6):47--56, 2003.

\bibitem{Lloyd2011}
D.~Lloyd and J.~Dykes.
\newblock {Human-centered approaches in geovisualization design: Investigating
  multiple methods through a long-term case study}.
\newblock {\em IEEE Trans. Visualization and Computer Graphics (TVCG) (Proc.
  InfoVis)}, 17(12):2498--2507, 2011.

\bibitem{McLachlan2008}
P.~McLachlan, T.~Munzner, E.~Koutsofios, and S.~North.
\newblock Liverac: Interactive visual exploration of system management
  time-series data.
\newblock In {\em Proceedings of the SIGCHI Conference on Human Factors in
  Computing Systems}, CHI '08, pages 1483--1492, New York, NY, USA, 2008. ACM.

\bibitem{Mendelow1991}
A.~Mendelow.
\newblock {(No Title}.
\newblock In {\em Proceedings of Second International Conference on Information
  Systems}, Cambridge, MA, USA, 1991.

\bibitem{Meyer2012}
M.~Meyer, M.~Sedlmair, and T.~Munzner.
\newblock {The Four-Level Nested Model Revisited: Blocks and Guidelines}.
\newblock In {\em Proceedings of the VisWeek Workshop Beyond Time and Errors:
  Novel Evaluation Methods for Information Visualization (BELIV)}. ACM Press,
  2012.

\bibitem{Meyer2013}
M.~Meyer, M.~Sedlmair, P.~S. Quinan, and T.~Munzner.
\newblock {The nested blocks and guidelines model}.
\newblock {\em Information Visualization}, 14, 2013.

\bibitem{Munzner2009}
T.~Munzner.
\newblock {A nested model for visualization design and validation.}
\newblock {\em IEEE Trans. Visualization and Computer Graphics (TVCG) (Proc.
  InfoVis)}, 15(6):921--8, 2009.

\bibitem{North2006}
C.~North.
\newblock {Toward measuring visualization insight}.
\newblock {\em IEEE Trans. Visualization and Computer Graphics (TVCG) (Proc.
  InfoVis)}, 26(3):6--9, 2006.

\bibitem{Quinan2016}
P.~S. Quinan and M.~Meyer.
\newblock {Visually Comparing Weather Features in Forecasts}.
\newblock {\em IEEE Trans. Visualization and Computer Graphics (TVCG) (Proc.
  InfoVis)}, 22(1):389--398, 2016.

\bibitem{Rindfleisch1997}
T.~C. Rindfleisch and T.~C. Rindfleisch.
\newblock {Privacy, information technology, and health care}.
\newblock {\em Communications of the ACM}, 40(8):92--100, 1997.

\bibitem{Safran2007}
C.~Safran, M.~Bloomrosen, W.~Hammond, S.~Labkoff, S.~Markel-Fox, P.~C. Tang,
  and D.~E. Detmer.
\newblock {Toward a national framework for the secondary use of health data: an
  American Medical Informatics Association White Paper}.
\newblock {\em Journal of the American Medical Informatics Association},
  14(1):1--9, 2007.

\bibitem{Sedlmair2011}
M.~Sedlmair, P.~Isemberg, D.~Baur, and A.~Butz.
\newblock {Information Visualization Evaluation in Large Companies : Challenges
  , Experiences and Recommendations}.
\newblock {\em Information Visualization Journal}, 10(3):248----266, 2011.

\bibitem{Sedlmair2010}
M.~Sedlmair, P.~Isenberg, D.~Baur, and A.~Butz.
\newblock Evaluating information visualization in large companies: Challenges,
  experiences and recommendations.
\newblock In {\em Proceedings of the CHI Workshop Beyond Time and Errors: Novel
  Evaluation Methods for Information Visualization (BELIV)}, 2010.

\bibitem{Sedlmair2012}
M.~Sedlmair, M.~Meyer, and T.~Munzner.
\newblock {Design Study Methodology: Reflections from the Trenches and the
  Stacks}.
\newblock {\em IEEE Trans. Visualization and Computer Graphics (TVCG) (Proc.
  InfoVis)}, 18(12):2431--2440, dec 2012.

\bibitem{Shneiderman2006}
B.~Shneiderman and C.~Plaisant.
\newblock {Strategies for evaluating information visualization tools}.
\newblock {\em Proceedings of the 2006 AVI workshop on BEyond time and errors
  novel evaluation methods for information visualization - BELIV '06}, page~1,
  2006.

\bibitem{Simon1981}
H.~A. Simon.
\newblock {\em {The sciences of the artificial(2nd ed.)}}.
\newblock MIT press, 1981.

\bibitem{Vicente1999}
K.~J. Vicente.
\newblock {\em {Cognitive Work Analysis: Toward Safe, Productive, and Healthy
  Computer-Based Work}}.
\newblock CRC Press, 1999.

\bibitem{Winters:2014}
K.~M. Winters, D.~Lach, and J.~B. Cushing.
\newblock Considerations for characterizing domain problems.
\newblock In {\em Proceedings of the Fifth Workshop on Beyond Time and Errors:
  Novel Evaluation Methods for Visualization}, BELIV '14, pages 16--22, New
  York, NY, USA, 2014. ACM.

\bibitem{Woolf2008}
S.~H. Woolf.
\newblock {The meaning of translational research and why it matters.}
\newblock {\em JAMA : the journal of the American Medical Association},
  299(2):211--213, 2008.

\bibitem{WorldHealthOrganization2013}
{World Health Organization}.
\newblock {Global Tuberculosis Report 2013}.
\newblock Technical report, 2013.

\end{thebibliography}
% ACM needs 'a single self-contained file'!
%\balancecolumns
%\appendix
%Appendix A
%\section{Headings in Appendices}
%The rules about hierarchical headings discussed above for
%the body of the article are different in the appendices.
%In the \textbf{appendix} environment, the command
%\textbf{section} is used to
%indicate the start of each Appendix, with alphabetic order
%designation (i.e. the first is A, the second B, etc.) and
%a title (if you include one).  So, if you need
%hierarchical structure
%\textit{within} an Appendix, start with \textbf{subsection} as the
%highest level. Here is an outline of the body of this
%document in Appendix-appropriate form:

%\balancecolumns % GM June 2007
% That's all folks!
\end{document}